\documentclass[twocolumn,preprintnumbers,superscriptaddress,amssymb,prl,aps,10pt]{revtex4-1}
\usepackage{graphicx}
\usepackage{dcolumn}
\usepackage{bm}
\usepackage{color}
\usepackage{amsmath,amsfonts,amssymb} 
\usepackage{graphicx}
\usepackage[english]{babel}
\usepackage{booktabs,longtable}
\usepackage{natbib}

\begin{document}

\author{Fernando~Ajejas}
\email{fernando.ajejas@cnrs-thales.fr}
\affiliation{Unit\'e Mixte de Physique, CNRS, Thales, Universit\'e Paris-Saclay, 91767, Palaiseau, France}
\author{Yanis~Sassi}
\affiliation{Unit\'e Mixte de Physique, CNRS, Thales, Universit\'e Paris-Saclay, 91767, Palaiseau, France}
\author{William~Legrand}
\affiliation{Unit\'e Mixte de Physique, CNRS, Thales, Universit\'e Paris-Saclay, 91767, Palaiseau, France}
\author{Sophie~Collin}
\affiliation{Unit\'e Mixte de Physique, CNRS, Thales, Universit\'e Paris-Saclay, 91767, Palaiseau, France}
\author{Andr\'e~Thiaville}
\affiliation{Laboratoire de Physique des Solides, Universit\'e Paris-Saclay, CNRS UMR 8502, 91405, Orsay Cedex, France}
\author{Jose~Pe\~na~Garcia}
\affiliation{Universit\'e Grenoble Alpes, CNRS, Institut N\'eel, F-38000 Grenoble, France}
\author{Stefania~Pizzini}
\affiliation{Universit\'e Grenoble Alpes, CNRS, Institut N\'eel, F-38000 Grenoble, France}
\author{Nicolas~Reyren}
\affiliation{Unit\'e Mixte de Physique, CNRS, Thales, Universit\'e Paris-Saclay, 91767, Palaiseau, France}
\author{Vincent~Cros}
\email{vincent.cros@cnrs-thales.fr}
\affiliation{Unit\'e Mixte de Physique, CNRS, Thales, Universit\'e Paris-Saclay, 91767, Palaiseau, France}
\author{Albert~Fert}
\affiliation{Unit\'e Mixte de Physique, CNRS, Thales, Universit\'e Paris-Saclay, 91767, Palaiseau, France}
\title{Element-selective modulation of interfacial Dzyaloshinskii-Moriya interaction in Pt$|$Co$|$Metal based multilayers.}

\date{\today}

\begin{abstract}
Despite a decade of research, the precise mechanisms occurring at interfaces underlying the Dzyaloshinskii-Moriya interaction (DMI), and thus the possibility of fine-tuning it, are not yet fully identified. In this study, we investigate the origin of the interfacial DMI, aiming at disentangling how independent are the interfaces around the ferromagnetic layer, and what are their relative contributions to the effective DMI amplitude. For this purpose, we have grown and investigated a large variety of systems with a common structure Pt$|$Co$|M$ with $M =$ Ni, Pd, Ru, Al, Al$|$Ta and MoSi. We explore the correlation between the effective interfacial DMI, and different intrinsic properties of metals, namely atomic number, electronegativity and work function difference at the Co$|M$ interfaces. We find a linear relationship between interfacial DMI and the work function difference between the two elements, hence relating the nature of this behavior to the interfacial potential gradient at the metallic interfaces. The understanding of the DMI mechanism is of utmost importance since it opens up the possibility of precisely engineering the magnetic and hence the spintronic properties for future devices.
\end{abstract}

\maketitle

In recent years, novel chiral magnetic textures such as chiral-N\'eel domain walls (DWs) \cite{Thiaville2012, Ryu2013, Emori2013}, spin spirals \cite{Ferriani2008} or skyrmions \cite{Fert2017} in thin films and multilayers have been at the core of many researches. The interest is two-fold due to their interesting fundamental physics and their potential as information carriers in the future generation of spintronic devices, e.g. storage, logic and neuromorphic devices \cite{Fert2017}. One of the key interactions involved in the stabilization of these chiral textures is the antisymmetric exchange, known as Dzyaloshinskii-Moriya interaction (DMI) \cite{Dzyaloshinskii1958, Moriya1960, Moriya1960a, Dzyaloshinskii1964, Fert1980}.
DMI is observed in systems with broken inversion symmetry and large spin orbit coupling (SOC) \cite{Fert1990}. These conditions are realized in thin film systems in which a ferromagnetic layer (FM) is sandwiched between a heavy metal (HM) and another metallic ($M$) or oxide layer. Despite the fact that the general conditions for its existence are identified, there are still open questions concerning the microscopic origin, amplitude or sign of the DMI. In multilayered systems with FM layers thinner than the exchange length, at least two interfaces have to be considered and the measured interfacial DMI corresponds to an effective one. This makes it difficult to disentangle if it results from adding independent contributions or from a combined effect of both interfaces. It is therefore of utmost importance to solve these questions in order to be able to tune DMI values on demand.

Beyond the Fert-Levy three-sites model between two atomic spins and a neighboring atom with a large SOC in diluted alloys \cite{Fert1980, Levy1990} and its extension to bilayer systems \cite{Fert1990}, first-principle calculations have been carried out for atomic monolayer (ML) samples, evaluating quantitatively the antisymmetric DM exchange interaction \cite{heide2008}. For few atomic layers, dissociating the contributions to the effective DMI ($D_{\rm eff}$) of each Pt ML adjacent to the Co MLs \cite{Zimmermann2014, Yang2015} showed diverse contributions up to several ML away from the interface. In addition, the role of the $3d$ orbital occupations and the hybridization with the spin-orbit active $5d$ states to modulate the amplitude of interfacial DMI was also observed \cite{Belabbes2016}. However, these calculations provide insight for specific systems, mostly epitaxial and single-layer, so that the large variety of experimental systems different from the ones described above, often non-epitaxial, using lighter elements \cite{Ajejas2017}, oxides \cite{Torrejon2014} or 2D materials such as graphene \cite{Ajejas2018} must still be explored.

The archetypal experimental multilayered system is composed of a thin FM layer deposited on a HM element, with Pt$|$Co being the most studied bilayer because of the large PMA  \cite{Cacia1988} and the large interfacial DMI \cite{Ma2018}. In most of the experimental studies (except, e.g., Ref.\,\cite{Corredor2017} under ultra-high vacuum conditions), a metal or oxide layer is grown on top of the FM layer, either to generate the next repetition in multilayers or as a protecting capping layer in simple trilayer systems. 
This layer has to be different from the bottom one, to guarantee the broken inversion symmetry and to avoid the cancellation of the interfacial DMI. An effective  DMI amplitude $D_{\rm eff}$ results from the combination of the DMI contributions of the two FM interfaces, either reinforcing or competing \cite{Moreau-Luchaire2016, Kim2016}. Our main objective is to understand how the interfaces operate, by focusing on the influence of the top Co$|M$ interface on the effective DMI amplitude $D_{\rm eff}$ in multilayers having structure  Pt$|$Co$|M$. This parallels the systematic theoretical work of Jia {\it et al}. \cite{Jia2020}, albeit without the periodic repetition of the trilayer, and with nanometer instead of 1 atomic ML thicknesses. The values of the effective interfacial energy $D_{\rm s}$ = $D_{\rm eff} \cdot t_{\rm Co}$, estimated by the asymmetric expansion of domain walls (DWs), are correlated with intrinsic material properties, namely the atomic number ($Z$), the electronegativity ($\chi$) and the work function ($\Phi$) already proposed by Park {\it et al.} \cite{Park2018}. These last two parameters are related to the interfacial potential gradient ($\nabla V$) at the metallic interfaces. Considering Pt$|$Co$|M$ with $M =$ Ni, Pd, Ru, Al, Al$|$Ta and MoSi allows us to explore different $d$ band filling and a large variety of intrinsic properties. Here, we deliver a first catalog of properties for the engineering of chiral textures \cite{Thiaville2012, Finco2021, Moreau-Luchaire2016, Legrand2020, Palermo2020}. 

The multilayer systems were grown by dc-magnetron sputtering at room temperature (RT) on thermally oxidized silicon substrates covered with a $280$-nm thick SiO$_2$ layer. The buffer layer is composed by 5\,nm Ta layer promoting good adherence on SiO$_{2}$ and inducing the $(1\,1\,1)$-oriented texture in the 8-nm thick Pt layer. This favors a strong uniaxial perpendicular anisotropy ($K_{\rm u} \sim 0.8-1.1\,{\rm MJ}/{\rm m}^3$ for Co thickness around 1-1.5 nm.) at the Pt$|$Co interface. A 3-nm thick Pt capping layer is deposited on top of all multilayers in order to prevent oxidation.
The magnetic characterization was performed by superconducting quantum interference device (SQUID) and alternative gradient field magnetometer (AGFM) magnetometry to measure the saturation magnetization ($M_{\rm s}$) and the effective anisotropy field ($\mu_0H_{K}=2K_{\rm eff}/M_s$).

In order to measure the value of the micromagnetic exchange constant $A$, Brillouin light scattering (BLS) measurements in the Damon-Eshbach (DE) geometry have been performed for samples with different Co thickness. In the DE geometry, the spin-wave frequencies are given, for ultrathin films, by \cite{Nembach2015}: 
\begin{equation} \label{eq1}
\begin{split}
f = f_{0} \pm f_{\rm DMI} \equiv \frac{\gamma \mu_{0}} {2\pi} \sqrt{[ \,H_{\rm ip} + Jk_{\rm sw} ^2+ \xi (k_{\rm sw}t)M_{\rm s}] \,} \cdot \\
\sqrt{[ \,H_{\rm ip} + Jk_{\rm sw} ^2- \xi (k_{\rm sw}t)M_{\rm s} - {H_{\rm K}}] \,} \pm \frac{\gamma} { \pi M_{\rm s}} D_{\rm  }k_{\rm sw} \ ,
\end{split}
\end{equation} 
where $\gamma$ is the Co gyromagnetic ratio, $\mu_{0}$ the magnetic permeability of vacuum, $H_{\rm ip}$ the in-plane external magnetic field, $J = 2A/(\mu_{0} M_{\rm s})$ is the spin wave stiffness, $k_{\rm sw} = 4\pi\sin(\theta)/\lambda$ is the spin-wave wavector, $\theta$ being the incidence angle of the laser wavelength $\lambda$ = 532 nm. Finally, $\xi$ accounts for the influence of dipolar interactions on the spin waves.

\begin{figure}[ht!]
\centering
\includegraphics[scale=0.5]{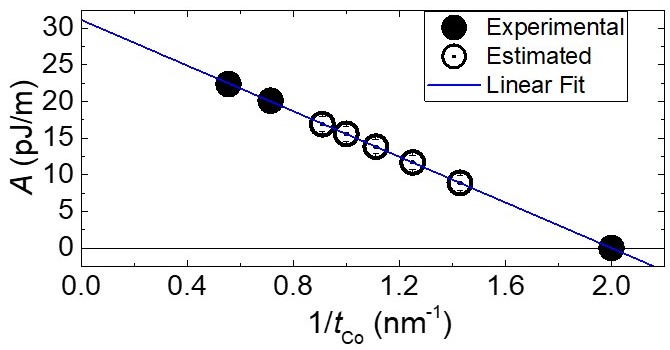}
\caption{Exchange constant $A$ as function of the inverse Co thickness $(1/t_{\rm Co})$. Filled circles are the BLS experimental $A$ measurements. Open circles are estimations from the linear fit for the Co thicknesses used in this paper.}
\label{figAvst}
\end{figure}

The exchange constant is obtained from the fit of $f_0$ \textit{vs} $k_{\rm sw}$ for Pt$|$Co(1.7)$|$Ru and Pt$|$Co(1.4)$|$Al trilayers, and from the thickness (0.5\,nm) where we have measured that the magnetization vanishes at RT. In Figure \ref{figAvst} we show the inverse of Co thickness dependence of the exchange constant for the samples measured (black filled dots). Assuming a phenomenological linear thickness variation, it is possible to extrapolate the values of $A$ for any thickness of interest (open circles). Interestingly, the extrapolated value at $1/t_{Co}=0$ (bulk Co thickness), shows a value of $31 \pm$1\,pJ/m in good agreement with the Co bulk value \cite{Coey2010}.
Table\,\ref{TableParam} summarizes the values of $A$ derived from extrapolation for the systems which have been used to derive the strength of DMI.

\begin{figure}[ht]
\centering
\includegraphics[scale=0.48]{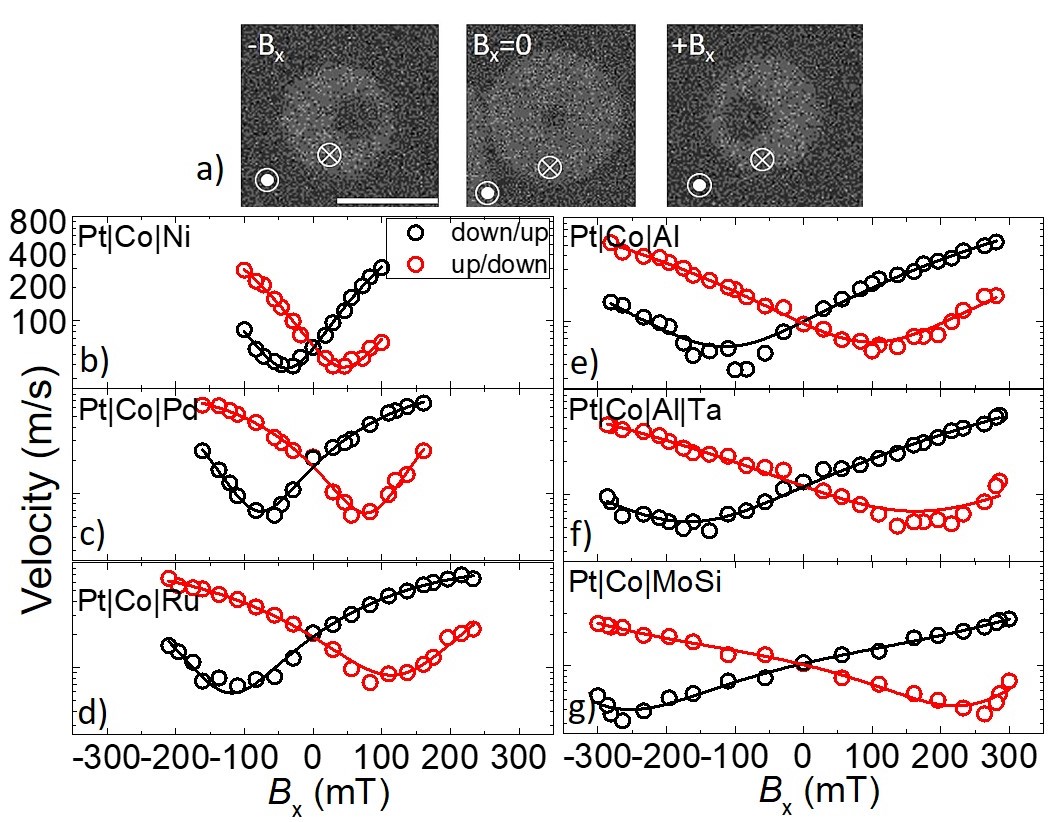}
\caption{ DMI determination using asymmetric domain expansion, as observed by Kerr microscopy. (a) Expansion of a N\'eel-type bubble domain in Pt$|$Co$|$Al trilayer for different external in-plane magnetic fields $B_x$ = -110, 0 and 110\,mT. Scale bar is 20\,$\mu$m. (b)-(g) Domain wall velocity {\it vs.} in-plane field $B_x$ for up/down and down/up DWs propagating in the x-direction. The driving $B_z$ = 250\,mT field induces a DW propagation in the flow precessional regime. Note that $t_{\rm Co}$ varies for the different materials (see Table \ref{TableParam}).
}
\label{figDWexp}
\end{figure}

The amplitude and sign of $D_{\rm eff}$ in the Pt$|$Co$|M$ samples is estimated from the asymmetric expansion of bubble domains measured by polar magneto-optical Kerr microscopy \cite {Je2013, Vanatka2015, yoshimura2016}. The domains are driven by an out-of-plane magnetic field pulses ($B_z$) of strength up to $500$\,mT in the presence of a static in-plane magnetic field ($B_x$). This large out-of-plane field is chosen to ensure that the DW dynamics are held in flow precessional regime \cite{PenaGarcia2021}, avoiding the smeared and non-symmetric DW expansion found in the creep regime \cite {Je2013, Vanatka2015, hrabec2014}. Details of the experimental procedure are described in previous works \cite{Vanatka2015, Pham2016, Ajejas2018}. 
The differential Kerr images in Figure\,\ref{figDWexp}(a), show the symmetric (asymmetric) expansion of a bubble domain driven by $B_z$ field pulses with $B_x = 0\,(B_ x\neq0)$ for the Pt$|$Co$|$Al system. The asymmetric expansion of the bubble domain is related to the presence of chiral N\'eel DWs, whose velocity depends on the relative direction of their internal magnetization and that of the in-plane magnetic field. A counter-clockwise (CCW) rotation of the magnetization, resulting from a negative $D_{\rm eff}$, is obtained for all the samples. The velocities of up/down and down/up DWs driven by a fixed out-of-plane field pulse, as a function of the $B_x$ in-plane field are shown in Figure\,\ref{figDWexp}(b-g). The in-plane field for which the DW velocity reaches a minimum is the one compensating the field $\mu_{0}H_{\rm DMI}=D_{\rm eff}/(\Delta M_{\rm s})$ that stabilizes the chiral N\'eel walls ($\Delta = \sqrt{A/K_{\rm eff}}$ is the domain wall parameter.). 
From this field value, the amplitude of the effective interfacial DMI energy $D_{\rm s}$ is estimated through:
\begin{equation}
D_{\rm s}=\mu_{0}H_{\rm DMI}M_{\rm s}t_{\rm Co}\Delta \ ,
\label{EqDsDW}
\end{equation} 
The results are shown in Table\,\ref{TableParam}, that includes all the magnetic parameters used in the calculations. 

For Pt$|$Co$|$Ni we find that $D_{\rm s}$ is $-0.21\pm 0.08$ pJ/m, a value that is within the range of indeed highly variable results in Co$|$Ni systems \cite{Lau2018, Yu2016}. For the Pt$|$Co$|$Pd system, $D_{\rm s}$ is $-0.80\pm 0.20$ pJ/m, which is a slightly higher value than those found in \cite{Guan2017, Koyama2018}. The system with Ru top layer presents a $D_{\rm s} = -1.04\pm 0.20$ pJ/m, this value is slightly smaller than the one determined by BLS measurements performed in Pt$|$Co(1.7)$|$Ru used for the estimation of $A$ described earlier.
For Pt$|$Co$|$MoSi, a system of interest for studies of the interplay between superconducting vortices in MoSi and the magnetic skyrmions in the ferromagnet \cite{Palermo2020}, we find $D_{\rm s}$ = $-0.93\pm 0.20$ pJ/m. Then, in Pt$|$Co$|$Al and Pt$|$Co$|$Al$|$Ta, we estimate $D_{\rm s} = -1.16$ and $ -1.46\pm 0.20$ pJ/m respectively. Interestingly, neither $M_{\rm s}$ nor $D_{\rm s}$ are identical in these two systems, suggesting that phenomena beyond the Fert-Levy three-site indirect exchange are at play. For the Pt$|$Co$|$Al, $D_{\rm s}$ is in agreement with that reported in Ref. \cite{Ajejas2017} where the samples were epitaxial. Note that in this case the results are in good agreement with those obtained by BLS for the same system (see Table \ref{TableParam}).

\begin{table*} [th!]
\caption{Measured magnetic properties of the different Pt$|$Co$|M$ trilayers: cobalt thickness $t_{\rm Co}$, metal thickness $t_{\rm M}$, exchange constant $A$, spontaneous magnetization $M_{s}$, effective anisotropy field $\mu_{0}H_{K}$, domain wall width parameter $\Delta$, DMI field $\mu_{0}H_{\rm DMI}$, effective interface DMI energy density $D_{\rm eff}$ extracted from the DMI field [Eq. (1)], effective interfacial DMI energy density $D_{\rm s}$, and effective interfacial DMI energy density $D_{\rm s}^{\rm BLS}$ measured by BLS. Negative DMI values in the table correspond to CCW-DW chirality. Error margins are the same for all sample combination. For clarity, they are only indicated in the first line, except for the error of $D_{\rm s}$, which is 0.20 $\rm pJm^{-1}$.}
\begin{center}
\begin{tabular}{ c c c c c c c c c c c c } 
\hline
Stacking & $t_{\rm Co}$& $t_{\rm M}$& $A$ & $M_{\rm s}$ & $\mu_{0}H_{\rm K}$ & $\Delta$ & $\mu_{0}H_{\rm DMI}$ & $D_{\rm eff}$ & $D_{\rm s}$ & $D_{\rm s}^{\rm BLS}$\\
& (nm) & (nm) & ($\rm pJm^{-1}$) & ($\rm MAm^{-1}$) & (T) & (nm) & (mT) & ($\rm mJm^{-2}$) & ($\rm pJm^{-1}$) & ($\rm pJm^{-1}$) \\ [0.2ex] 
\hline\hline
Pt$|$Co$|$Ni & 0.7 & 0.4 & $8.9\pm 0.5$ & $0.97\pm 0.05$ & $0.46\pm 0.05$ & $6.3\pm 0.5$ & $50\pm 10$ & $-0.31\pm 0.10$ & $-0.21\pm 0.08$ & -- \\
\hline
Pt$|$Co$|$Pd & 0.9 & 1.0 & 13.8 & 1.80 & 0.35 & 6.6 & $75$ & $-0.89$ & $-0.8\pm 0.20$ & -- \\
\hline
Pt$|$Co$|$Ru & 1.1 & 1.4 & 17.3 & 1.10 & 0.52 & 7.8 & 110 & $-0.95$ & $-1.04$ & $-1.27\pm 0.03$ \\ 
\hline
Pt$|$Co$|$Al & 1.0 & 1.4 & 15.6 & 1.10 & 0.31 & 9.6 & 110 & $-1.16$ & $-1.16$ & $-1.07\pm 0.09$ \\
\hline
Pt$|$Co$|$Al$|$Ta & 1.0 & 1.4$|$3.0 & 15.6 & 0.96 & 0.38 &9.3& 165 & $-1.46$ & $-1.46$ & -- \\
\hline
Pt$|$Co$|$MoSi & 0.8 & 1.4 & 10.4 & 0.87 & 0.90 & 5.2 & 260 & $-1.16$ & $-0.93$ & -- \\
\hline
\end{tabular}
\end{center}
\label{TableParam}
\end{table*}

Considering the body of knowledge gained by the first principles calculations of DMI vs sample composition and architecture, we investigate experimentally the possible correlation between the estimated $D_{\rm s}$ values and the metal properties such as atomic number ($Z$), the electronegativity ($\chi$) or the work function ($\Phi$). The first one is related to the strength of the SOC and the latter two are related to the expected interfacial $\nabla V$ which gives rise to Rashba or interfacial DMI effects \cite{Heide2006, Bihlmayer2006}. In order to increase the palette of material systems that can be analyzed, we have added to our experimental data set some of our previously reported experimental $D_{\rm eff}$ values, namely, Pt$|$Co$|M$, M$=$Pt, Ir, Cu \cite{Ajejas2017}, Gd \cite{Pham2016} and Pt$|$Co$|$Graphene \cite{Ajejas2018} (open circles in following figures). Note that these additional measurements were performed using the same MOKE setup.

\begin{figure}[ht]
\centering
\includegraphics[scale=0.45]{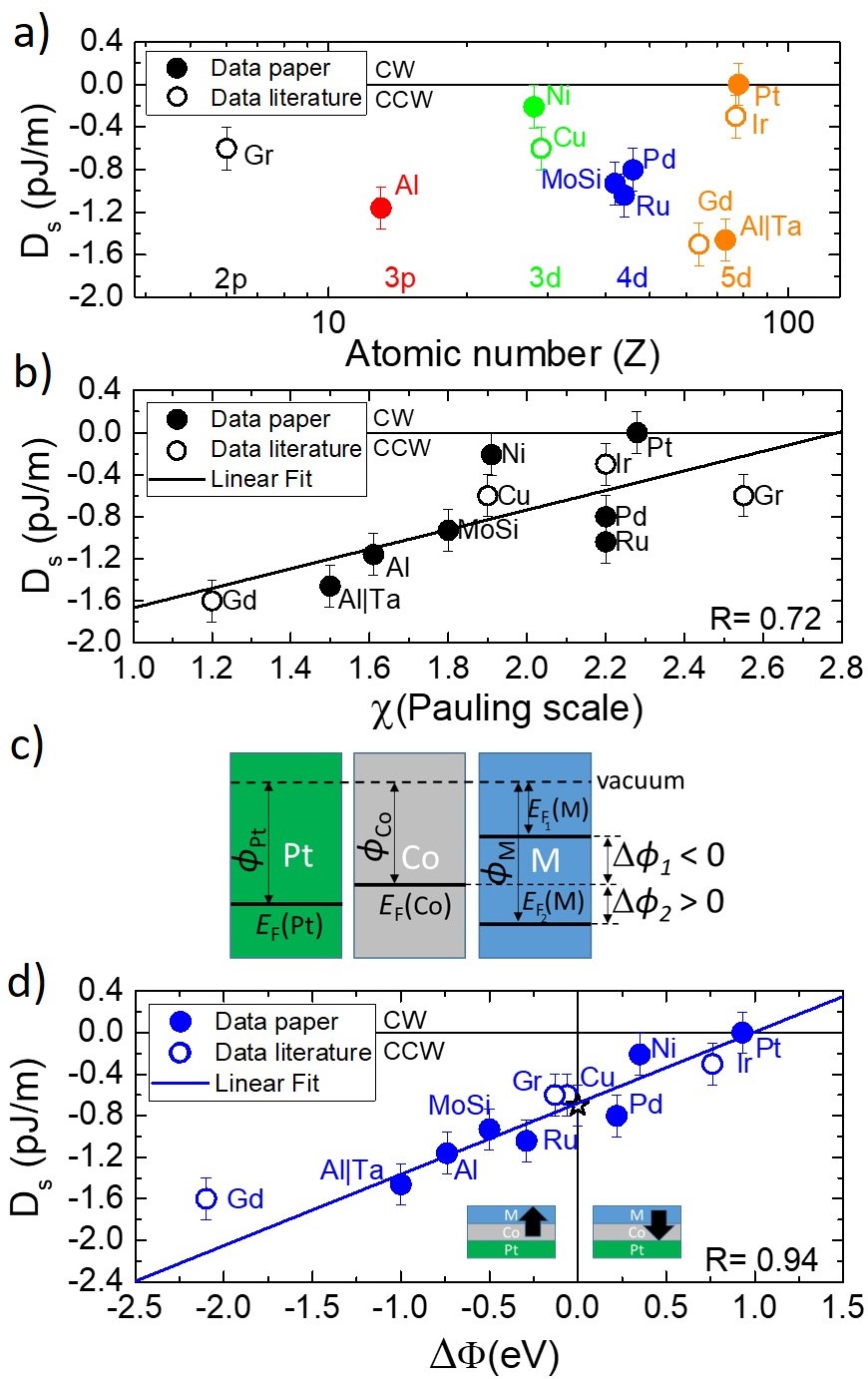}
\caption{Correlation of parameters characterizing the Co$|$M interfaces with the effective DMI, $D_{s}$. (a) Atomic Number ($Z$) dependence; (b) Electronegativity ($\chi$); and (d) work function difference between Co and M layer ($\Delta\Phi$), for Pt$|$Co$|$M samples with different M layers. (c) Scheme representing the trilayer structure with Fermi levels, $\Phi$ and $\Delta\Phi$. $\Delta\Phi$ values are calculated from the literature. Open circles are extracted from Ref.\,\cite{Ajejas2017, Pham2016, Ajejas2018}.}
\label{figDsvsX}
\end{figure}

We first analyze as first parameter the strength of the SOC. It is known that the SOC increases with the atomic number ($\propto Z^4$ for isolated atoms). However, a precise quantitative dependence with $Z$ is not yet fully established in metallic multilayers. In particular, for the heaviest elements, even though the number of electrons that interacts with the nucleus increases, it must be considered that the core and the outer electrons are not exposed to the same effective fields. For our case of interest, in the indirect exchange picture, the most relevant electrons are the outer ones and a SOC strength proportional to $Z^2$ is expected \cite{Landau1991, Heide2009}. 
In Figure\,\ref{figDsvsX}(a), we present the evolution of the measured $D_{\rm s}$ as a function of the atomic number $Z$. Interestingly, we find that the experimental $D_{\rm s}$ are sorted out in different groups or bunches depending on their outermost level, i.e. $2p$, $3p$, $3d$, $4d$ and $5d$.
Because the plotted $D_{\rm s}$ is an effective value including the Pt$|$Co interface, it is however difficult to evidence a power law relation with $Z$. Thus this parameter can be hardly considered as a simple parameter for the control of $D_{\rm s}$.

The second parameter, as $D_{\rm s}$ is expected to depend upon a charge transfer effect at the interfaces \cite{Jia2020}, is the interfacial $\nabla V$ resulting from the broken inversion symmetry. This parameter can be related to the electronegativity $\chi$ in Pauling scale \cite{Pauling1932} which is the ability of an atom to attract electrons when it combines with another atom in a chemical bond.
The dependence of $D_{\rm s}$ on $\chi$ is displayed in Figure\,\ref{figDsvsX}(b). A linear correlation is found, however with a relativity large dispersion that can be characterized by the Pearson's factor $R$, being equal to 0.72 in this case. We notice that this linear behavior is in good agreement with Jia {\it et al.} calculations \cite{Jia2020}, which were however obtained in ultra-thin systems, typically Pt(1\,ML)$|$Co(1\,ML)$| M$(1\,ML). They obtain a variation in amplitude of $5$\,pJ/m for the theoretical $D_{\rm s}$ calculations depending on the top $M$ element \cite{Jia2020}, while experimentally we find a total amplitude variation of $1.6$\,pJ/m between Pt and Gd. Since we are using multilayers with much thicker individual layers, we can anticipate that the properties of each interface might change significantly compared to these predictions. This difference indeed questions how independent the two interfaces can be considered and what is their mutual influence. 

The interfacial $\nabla V$ can also be related with the work function that represents the minimum energy needed to remove an electron from the solid to the vacuum \cite{Kittel1996}, determining the band alignment. For thick enough magnetic layers, e.g., 4-5 MLs, the two Pt$|$Co and Co$|$$M$ interfaces might be considered relatively independent, having each a Fermi level $E_{\rm F}$ that is specific to each element at a given interfaces [Figure\,\ref{figDsvsX}(c)].
Since in all systems the bottom Pt$|$Co interface is the same, we consider only the work function difference between Co and $M$: $\Delta\Phi = \Phi_M - \Phi_{\rm Co}$, and we expect $\nabla V\propto \Delta\Phi$. The same effect in terms of $\nabla V$ is expected at the bottom interface.
In Figure\,\ref{figDsvsX}(d), we present the evolution of the experimental $D_{\rm s}$ as a function of $\Delta\Phi$. Work function values are collected from the literature \cite{Michaelson1977, Skriver1992, Giovannetti2008}.
A linear relationship with a much better Pearson's parameter, $R = 0.94$, than for $\chi$ is found, indicating a better correlation. Only Gd is found slightly out of the line. The case of Gd is more complicated due to the $4f$ electrons and its magnetic behavior
. The insets in the graph represent schematically the trilayer structure, with an arrow indicating the direction of the interfacial $\nabla V$ due to $\Delta\Phi$. Depending on the sign of $\nabla V$, the interfacial DMI generated at the Co$|$M interfaces enhances ($\Delta\Phi < 0$) or decreases ($\Delta\Phi > 0$) the total $D_{\rm eff}$. 
Note that the experimental $D_{\rm s}$ value obtained in \cite {Corredor2017} for Pt$|$Co ($-0.7$\,pJ/m) [black star in Figure\,\ref{figDsvsX}(d)], have also added to the plot. Interestingly, it can be seen that this $D_{\rm s}$ value is close to one measured for Pt$|$Co$|$Cu ($D_{\rm s} = -0.6$\,pJ/m \cite {Ajejas2017}) for which the work function difference for this interface is only $6$\,meV. From this, we then conclude that for Pt$|$Co$|$Cu the effective DMI is originating from the Pt$|$Co bottom interface with almost no additive or subtractive effect from the Co$|$Cu interface. 

Finally, as it has been proposed for 2D electron gas systems \cite{Kim2013}, the interfacial DMI present in our metallic interfaces might be correlated to the Rashba parameter. In fact, the Rashba coefficient ($\alpha_{\rm R}$) \cite{Manchon2015} expressed through $D = 2k_{\rm R}A$, with $k_{R} = 2 \alpha_R m_e/ \hbar^2$ is generated by $\nabla V$ and then, is proportional to $\Delta\Phi$.

In conclusion, we have determined the value of $D_{\rm eff}$ in a series of sputtered Pt$|$Co$|M$ multilayers with $M =$ Ni, Pd, Ru, Al, Al$|$Ta and MoSi. The effective interfacial energy $D_{\rm s}$ is extracted from the asymmetric expansion of magnetic domains in the presence of an in-plane magnetic field. Its amplitude strongly depends on the nature of the top Co$|M$ interface. The best correlation is found with the work function difference between Co and the top M layer. The interfacial potential gradient at the Co$|M$ interface plays the crucial role pointing directly to an additive or subtractive behavior of the interfaces. Nevertheless in this $t_{\rm Co}$ range ($\approx 1\,$nm) an indirect modification of the Pt$|$Co bottom interface due to charge transfer may also play a role. The experimental observation of a strong correlation between $\Delta\Phi$ and $D_{\rm s}$ provides an efficient tool to obtain desired values of the effective DMI for the design of spin textures.

\begin{acknowledgments}
We acknowledge Prof. Stefan Blügel, Dr. Hongying Jia and Dr. Markus Hoffmann for fruitful discussions and suggestions. 
This work has been supported by DARPA TEE program grant $(MIPR HR-0011831554)$, ANR grant TOPSKY (ANR$-17-$CE$24-0025$), the FLAG-ERA SographMEM (ANR-15-GRFL-0005) and the Horizon2020 Framework Program of the European Commission, under FETProactive Grant agreement No. 824123 (SKYTOP) (H$2020$ FET Proactive 824123). 
J.P.G. acknowledges the European Union’s Horizon 2020 research and innovation program under Marie Sklodowska-Curie Grant Agreement No. 754303 and the support from the Laboratoire d'excellence LANEF in Grenoble (ANR-10-LABX-0051).
In addition to PhOM and EOE departments from Univ. Paris-Saclay, CNRS INP, the Sesame Ile de France IMAGeSPIN project (n°EX039175) and the French National Research Agency as part of the "Investissements d'Avenir" program by LABEX NanoSaclay (ANR10 LABX0035, BLS@PSay and SPiCY) are acknowledged for the BLS equipment.
\end{acknowledgments}

\bibliography{Biblio}

\end{document}